\documentclass[aps,twocolumn,showpacs,nofootinbib,amssymb]{revtex4}
\usepackage{graphicx,epsfig}
\newcommand{\nn}{\nonumber}
\newcommand{\be}{\begin{equation}}
\newcommand{\ee}{\end{equation}}
\newcommand{\bea}{\begin{eqnarray}}
\newcommand{\eea}{\end{eqnarray}}

\def\als{\alpha_{\rm s}}
\def\siml{{\ \lower-1.2pt\vbox{\hbox{\rlap{$<$}\lower6pt\vbox{\hbox{$\sim$}}}}\ }} 

\newcommand{\MS}{\overline{\rm MS}}

\begin{document}
\preprint{UAB-FT-XXX}

\title{\boldmath 
The 
static potential in ${\cal N}$=4 supersymmetric Yang-Mills 
at weak coupling
\unboldmath}
\author{Antonio Pineda}
\affiliation{Grup de F\'\i sica Te\`orica and IFAE, Universitat
Aut\`onoma de Barcelona, E-08193 Bellaterra, Barcelona, Spain}

\date{\today}

\begin{abstract}
\noindent
We compute the static potential associated to the locally 1/2 BPS 
Wilson loop in ${\cal N}$=4 supersymmetric Yang-Mills theory with ${\cal O}(\lambda^2/r)$ 
accuracy. We also resum the leading logarithms, of ${\cal O}(\lambda^{n+1}\ln^n\lambda/r)$, 
and show the structure of the renormalization group 
equation at next-to-leading order in the multipole expansion.
In order to obtain these results it is crucial the use of an effective theory for the 
ultrasoft degrees of freedom. We develop this theory up to 
next-to-leading order in the multipole expansion. Using the same formalism we 
also compute the leading logarithms, of ${\cal O}(\lambda^{n+3}\ln^n\lambda/r)$,
of the static potential associated to an ordinary Wilson loop in the same theory.  
\end{abstract}
\pacs{ 11.30.Pb, 11.10.Hi, 12.39.Hg} 
\maketitle


There is a huge interest in the study of the Yang-Mills theory 
with ${\cal N}=4$ supersymmetry in four dimensions. One of the 
reasons is the conjectured existence of a 
correspondence between the ${\cal N}$=4 supersymmetric Yang-Mills 
and the type IIA superstring theory on an AdS$_5\times S^5$ background
\cite{AdS}. This duality is known as the AdS/CFT correspondence and
has importance consequences, since it allows to compute the strong 't Hooft 
coupling limit ($\lambda \equiv \frac{N_cg^2}{4\pi}$) 
of certain correlators in the ${\cal N}$=4 supersymmetric Yang-Mills 
theory for large $N_c$. This is so because this limit becomes equivalent to 
the classical limit on the string theory side for which computational 
techniques exist. 

Checks of this conjecture are difficult to obtain, since the only quantitative 
approach to gauge theories is based on computations at weak coupling. 
In some cases it is possible to check the conjecture with the result at weak 
coupling. This usually happens when non-renormalization theorems exists 
that permit to perform the computation on the perturbative side exactly. 

In other cases no such checks exist and, usually, one can only study both the weak and 
strong coupling limit with increasing degree of accuracy hoping to gain 
further input on how the extrapolation from weak to strong coupling takes 
place. In this paper we concentrate in one of such correlators, the locally 
1/2 BPS static Wilson loop, 
or, more specifically, the large time limit of its logarithm. Its strong 't Hooft 
coupling limit for large $N_c$ has been computed using the AdS/CFT correspondence in 
\cite{Rey:1998ik,Vsstrong}. On the other hand the question whether the understanding 
of the infrared structure of the static potential in the weak coupling regime may 
shed some light on the AdS/CFT correspondence has been addressed in Refs. 
\cite{Erickson:1999qv,Erickson:2000af}. 
In these references some infrared divergences were found, which allowed the authors to obtain  
the leading logarithmic correction to the tree level result. These infrared divergences 
have a similar origin to those found in the QCD static potential at weak coupling \cite{Appelquist:1977es}, 
which, however, in this case first appear at three loops. They are due to the existence of 
degrees of freedom with energy of ${\cal O}(\lambda/r)$, which we will call ultrasoft in what follows. 
This scale is much smaller than the soft scale $\sim 1/r$ at weak coupling. In any case, it is somewhat 
surprising that, after so much work, the static potential has not even been computed with next-to-leading-order 
(NLO), ie. ${\cal O}(\lambda^2/r)$, accuracy yet. The reason is that, at this order, 
ultrasoft effects enter into the game and an infinite resummation of diagrams is needed 
in order to obtain the desired accuracy. However, this problem can be revisited 
from an effective field theory perspective, as it has already been done for the QCD case  
\cite{Brambilla:1999qa}. As we will see, by doing so, the problem trivializes and we will able to 
compute the static potential with ${\cal O}(\lambda^2/r)$ accuracy. The use of effective 
theories will also allow us to write renormalization group equations for the 
static potential. By solving them we will also obtain the static potential with leading-log (LL), 
ie. ${\cal O}(\lambda^{n+1}\ln^n\lambda/r)$, accuracy.

We will also use this example to give full details of how the
 factorization between the soft and ultrasoft scale takes place in 
the static potential for a specific computation 
including finite pieces. In the case of QCD this factorization 
takes places at three loop and the full computation does not exist yet. 

Finally, using the same formalism we will also compute the static potential 
associated to an ordinary Wilson loop with NNLL accuracy up to the two-loop 
matching condition.

\section{The static singlet energy}
The ${\cal N}=4$ SUSY Lagrangian reads
\bea
{\cal L}_{{\cal N}=4}&=& - {1\over 4} F_{\mu \nu}^{a} F^{\mu \nu \, a}
+
{1\over 2}\sum_{i=1}^6 (D_{\mu} \Phi_i)^a (D^{\mu}\Phi_i)^a
\nn
\\
&&
-{i\over 2}{\bar \Psi}^a \gamma_{\mu} (D^{\mu}\Psi)^a
+
\cdots
\,.
\label{LN4}
\eea
$\Phi$ and $\Psi$ represent a scalar and majorana particle respectively. The adjoint 
covariant derivative reads: $D_{\mu}(\cdot)^a=\partial_{\mu}(\cdot)^a-gf^{abc}A^b_{\mu}(\cdot)^c$,  
$i$ range from 1 to 6, and the dots represent interactions between the
scalars and the majorana particles or self-interactions between the
scalars.  

We now take the locally 1/2 BPS Wilson loop (for a motivation of this 
definition see \cite{Drukker:1999zq})
\be
W_C=\frac{1}{N_c}{\rm Tr} {\cal P} e^{-ig\int_C d\tau(A_{\mu}{\dot x}^{\mu}+\Phi_ n|{\dot x}|)}
\,,
\ee
where $\Phi_n \equiv \Phi\cdot {\hat n}$, $n$ is a six-dimensional vector with $n^2=1$, 
and $C$ represents the path followed by the Wilson loop. 
In our case we consider a static Wilson loop and its associated singlet static energy
\be
\label{Es}
E_s(r)=\lim_{T \rightarrow \infty}\frac{i}{T}\ln \langle W_\Box \rangle
\,,
\ee
where $W_\Box$ is the rectangular Wilson loop with edges $x_1 = (T/2,{\bf r}/2)$, $x_2 = (T/2,-{\bf r}/2)$, 
$y_1 = (-T/2,{\bf r}/2)$ and  $y_2 = (-T/2,-{\bf r}/2)$. 
The symbol $\langle ~~ \rangle$ means the average over the massless fields.

In order to describe the static Wilson loop at the dynamical level we consider 
the Lagrangian (\ref{LN4}) plus static sources in the appropriate representation 
\be
\label{LNRQCD}
{\cal L}={\cal L}_{{\cal N}=4}+{\cal L}_{stat.}
\,,
\ee
where
\be
{\cal L}_{stat.}=\psi^{\dagger}(i\partial_0-gA_0-g\Phi_n)\psi+
\chi_c^{\dagger}(i\partial_0+gA_0^{T}-g\Phi^{T}_n)\chi_c
\,.
\ee
$\psi$ and $\chi_c$ (the conjugated field of $\chi$) correspond to static sources 
in the fundamental and anti-fundamental 
representation respectively. The case with only one static source has been 
considered in Ref. \cite{Gomis:2006sb}.

With this Lagrangian the static singlet energy can be obtained from the following Green 
function
\begin{eqnarray}
I &\equiv& \langle 0 \vert  \chi^\dagger(x_2) W(x_2,x_1) \psi(x_1) \nonumber\\ 
& & \qquad \times \psi^\dagger(y_1)W(y_1,y_2) \chi(y_2) \vert 0 \rangle. 
\nn
\\
&=& \delta^3({\bf x}_1 - {\bf y}_1) 
\delta^3({\bf x}_2 - {\bf y}_2) \langle W_\Box \rangle ,
\label{vsnrqcd}
\eea
where $W(x_2,x_1)=W_C$ for a straight path $C$ with initial and final points 
$x_1$ and $x_2$ respectively. 

\section{Effective theory}

The use of effective field theories allows us for a efficient description 
of the infrared structure of the static potential at weak coupling. 
This has been shown to be so for the QCD static potential 
\cite{Brambilla:1999qa,Kniehl:1999ud}. In that case the effective theory 
was pNRQCD \cite{Pineda:1997bj}. For a review see \cite{Brambilla:2004jw}. 
Here we would like to construct the ${\cal N}=4$ supersymmetric version 
of pNRQCD in the static limit. We will do so at next-to-leading order 
in the multipole expansion. As we will see the main difference is 
the existence of massless scalars. On the other hand the heavy quark 
and antiquark rearrange in a singlet or octet configuration under 
(ultrasoft) gauge transformations as in QCD. The simplification arises from the 
existence of two disparate scales: the soft $\sim 1/r$ and the ultrasoft 
$\sim \lambda/r$ scales, and that we only aim for a description of the 
dynamics at energies of order $\sim \lambda/r$. Therefore, degrees of 
freedom with energy $\sim 1/r$ can be integrated out and one can factorize 
the physics associated to each scale. Note that we live in the opposite 
limit to strong coupling. For us $E_s(r) \ll 1/r$, whereas in the strong 
coupling limit $E_s(r) \gg 1/r$.

Integrating out the soft scale, $1/r$, from (\ref{LNRQCD}) we are left 
with an effective theory where only ultrasoft degrees of freedom remain 
dynamical. The surviving fields are the $\psi$-$\bar \chi$ states  (with ultrasoft energy) 
and the massless ultrasoft gluons, scalars and fermions.  The $\psi$-$\bar \chi$ states can be decomposed 
into a singlet (S) and an octet (O) under colour transformation. The relative coordinate 
${\bf r}= {\bf x}_1-{\bf x}_2$, whose typical size is the inverse of the soft scale, 
is explicit and can be considered as small with respect to the remaining (ultrasoft) 
dynamical lengths in the system. Hence the massless fields
can be systematically expanded in $\bf r$ (multipole expansion), and the effective 
Lagrangian is constructed order by order in 
${\bf r}$. As a typical feature of an effective theory, all the non-analytic 
behaviour in ${\bf r}$ is encoded in the matching coefficients, which can be interpreted as potential-like terms. 

In order to have the proper free-field normalization in the colour space we define 
\begin{equation}
{\rm S} \equiv { 1\!\!{\rm l}_c \over \sqrt{N_c}} S \quad \quad {\rm O} \equiv  { T^a \over \sqrt{T_F}}O^a, 
\label{norm}
\end{equation}
where $T_F=1/2$.  

The effective Lagrangian density that can be constructed with these 
fields and that is compatible with the symmetries of Eq. (\ref{LNRQCD}) is given at the next-to-leading 
order in the multipole expansion by:
\begin{eqnarray}
& & {\cal L}_{\rm US} =
{\rm Tr} \Biggl\{ {\rm S}^\dagger \left( i\partial_0  - V_s(r) + \dots  \right) {\rm S} 
\nonumber \\
& &\qquad + {\rm O}^\dagger \left( iD_0 - V_o(r) + \dots  \right) {\rm O} \Biggr\}
\nonumber\\
& &\qquad
-2gV_{\Phi}(r){\rm Tr}\left\{{\rm S}^{\dagger}\Phi_n {\rm O}+{\rm S}\Phi_n 
{\rm O}^{\dagger}\right\}
\nonumber\\
& &\qquad 
-gV_{\Phi_O}(r){\rm Tr}\left\{{\rm O}^{\dagger}\left\{\Phi_n, {\rm O}\right\}\right\}
\nonumber\\
& &\qquad 
+ g V_A (r) {\rm Tr} \left\{  {\rm O}^\dagger {\bf r} \cdot {\bf E} \,{\rm S}
+ {\rm S}^\dagger {\bf r} \cdot {\bf E} \,{\rm O} \right\} 
\nonumber \\
& &\qquad  + g {V_B (r) \over 2} {\rm Tr} \left\{  {\rm O}^\dagger \left\{{\bf r} \cdot {\bf E} , {\rm O}\right\}\right\} 
\nonumber \\
& &\qquad 
- g {V_C (r) \over 2} {\rm Tr} \left\{  {\rm O}^\dagger \left[{\bf r} \cdot ({\bf D}\Phi_n) , {\rm O}\right]\right\} ,  
\label{pnrqcd0}
\end{eqnarray}
where ${\bf R} \equiv ({\bf x}_1+{\bf x}_2)/2$, 
${\rm S} = {\rm S}({\bf r},{\bf R},t)$ and 
${\rm O} = {\rm O}({\bf r},{\bf R},t)$ are the singlet and octet wave functions respectively. 
All the gluon and scalar fields, as well as the derivative of them, 
in Eq. (\ref {pnrqcd0}) are evaluated 
in ${\bf R}$ and $t$. In particular 
${\bf F}^{i0}\equiv {\bf E} \equiv {\bf E}({\bf R},t)$ and 
$iD_0 {\rm O} \equiv i \partial_0 {\rm O} - g [A_0({\bf R},t),{\rm O}]$. 
$V_X(r)$ are the matching coefficients of the effective Lagrangian. They are determined 
by matching the effective and the underlying theory at a scale $\nu$ smaller 
than $1/r$ and larger than the ultrasoft scales. Since we are at weak coupling,
 the matching can be done perturbatively. 
 At the lowest order in the coupling 
constant we get $\alpha_{V_s} = \alpha_{V_o} = \alpha_{\rm s}\equiv \frac{g^2}{4\pi}$, 
$V_A=V_B=V_{\Phi}=V_{\Phi_O}=1$, where 
we have defined ($C_A=N_c$, $C_F=(N_c^2-1)/(2N_c)$) 
\begin{eqnarray}
\label{Vstree}
V_s(r) &\equiv&  - 2C_F {\alpha_{V_s}(r) \over r}, 
\label{defpot}\\ 
V_o(r) &\equiv&  2\left({C_A\over 2} -C_F\right) {\alpha_{V_o}(r) \over r},
\nonumber
\end{eqnarray}
which correspond to the singlet and octet heavy $\psi$-$\bar \chi$ static potential respectively. 

Note that we distinguish between the singlet static energy $E_s(r)$ and the singlet static potential 
$V_s(r)$. The reason for that has been discussed in detail in Ref. \cite{Brambilla:1999qa}. 
$V_s(r)$ corresponds to the matching coefficient that appear in the effective theory and it is 
the proper object to be introduced in a Schroedinger equation in case we were working with 
particles with large but finite mass. On the other hand $E_s(r)$ directly corresponds to Eq. (\ref{Es}) 
and represents the energy of two static particles (one could do a similar distinction for the 
octet static potential and energy, which in the QCD case corresponds to the hybrid energy).

Charge conjugation proves to be a very useful symmetry to eliminate operators in the 
effective Lagrangian. The effective Lagrangian is invariant under charge conjugation 
plus particle $ \leftrightarrow $ antiparticle exchange. In particular singlet, octet, gluon and scalar 
fields transform as:  
\bea 
&&
{\rm S}(t,{\bf r}, {\bf R}) \rightarrow \sigma^2 {\rm S}(t,-{\bf r}, {\bf
  R})^T  \sigma^2 , 
\\
&& 
\nn 
{\rm O}(t,{\bf r}, {\bf R}) \rightarrow \sigma^2  {\rm O}(t,-{\bf r}, {\bf
  R})^T  \sigma^2 , 
\\
&& 
\nn
A_\mu(t,{\bf R}) \rightarrow - A_\mu(t,{\bf R})^T,
\quad 
\Phi_n(t,{\bf R}) \rightarrow  \Phi_n (t,{\bf R})^T.
\eea
Asking for invariance under the above transformations the following operators are 
eliminated:
\be
\delta {\cal L}=-gV_D{\rm Tr}\left\{{\rm O}^{\dagger}[\Phi_n,{\rm O}]\right\}
\,,
\ee
\be
\delta {\cal L}
=
-gV_E{\bf r}\left({\rm Tr}\left\{{\rm S}^{\dagger}({\bf D}\Phi_n) {\rm O}\right\}+h.c.\right)
\,,
\ee
\be
\delta {\cal L}
=
-gV_F{\bf r}{\rm Tr}\left\{{\rm O}^{\dagger}\left\{({\bf D}\Phi_n),{\rm O}\right\}\right\}
\,.
\ee
One can also easily check perturbatively up to order $\als^2$ that the matching coefficient 
of these operators are zero. Therefore, 
they do not affect the results obtained in this paper for the singlet static potential and energy. 

We have to mention that we have not explored all the constraints supersymmetry (or any other remaining underlying 
symmetry) may impose on the 
matching coefficients of the effective Lagrangian and, accordingly, on the structure 
of the renormalization group equation. This would need a dedicated study that goes beyond the aim of this
work. In the case of QCD some analysis have been performed for the underlying Poincare symmetry of pNRQCD in 
Ref. \cite{Brambilla:2003nt}.

Finally, let us note that the octet singlet potential is $1/N_c^2$ suppressed compared with 
the singlet potential. Therefore, it can be neglected in the large $N_c$ limit. 

\section{RG}

The renormalization group equation of the singlet static potential 
has the following structure
\be
\nu \frac{d}{d\nu}V_s=\gamma_1(V_s-V_o)+\gamma_3r^2(V_s-V_o)^3+\cdots
\,,
\ee
where $\nu$ is the factorization scale, and 
the dots refer to higher orders in the multipole expansion. 
This structure follows from the multipole expansion and the mass gap between 
the octet and singlet static potential. 
The anomalous dimensions $\gamma_i$ have structure themselves. 
$\gamma_1$ can be computed as an expansion in 
 $\als$ (note that some powers of $\sqrt{\als}$ 
 get associated a power of $V_{\Phi}$ or $V_{\Phi_O}$). For $\gamma_3$ we also 
 have dependence on $V_A^2$ (but in a very specific way):
\be
\gamma_3=V_{A}^2(2C_F\als) F(\als,V_{\Phi},V_{\Phi_O})
\ee 
For both anomalous dimensions we can easily obtain the lowest non-trivial 
 contribution. They read
\be
\gamma_1=2\frac{2C_F\als}{\pi}V_{\Phi}^2+{\cal O}(\als^2)
\,,
\ee
\be
\label{g3}
\gamma_3=\frac{2}{3}\frac{2C_F\als}{\pi}V_{A}^2+{\cal O}(\als^2)
\,.
\ee
They come from the computation of the ultraviolet behavior of the 
diagram in Fig. \ref{figus} with the $V_{\Phi}$ or $V_A$ vertex. Note that 
the computation of $\gamma_1$ 
is explicitly gauge invariant in the effective theory and 
only needs the computation of one diagram. Therefore, it does not need of the 
delicate cancellation between infrared divergences that takes 
place at the soft scale, in particular, in the evaluation of the leading 
logarithm made in Ref. \cite{Erickson:2000af}, upon which the contribution of each 
particular diagram is gauge dependent. 
Quite remarkably, the leading order contribution to 
$\gamma_3$ is the same as the one one would obtain in pure QCD, since 
it is only due to the chromoelectric field. On the other hand in 
QCD $\gamma_1=0$ \cite{Pineda:2000gz}.
\begin{figure}[htb]
\makebox[0.0cm]{\phantom b}
\epsfxsize=5truecm \epsfbox{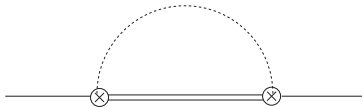}
\caption{\it One loop contribution to the singlet propagator. The dotted line 
represents the scalar field.}
\label{figus}
\end{figure}

The large logarithms are resummed by running $\nu$ 
from $ \sim 1/r$ to $\sim \lambda/r$ in the solution 
of the renormalization group equation. In our case,  
using that $V_{\Phi}=1$ at LL, we can solve the renormalization group 
equation for $V_s$ (for simplicity 
in the rest of this section we take the large $N_c$ limit):
\be
\nu \frac{d}{d\nu }V_s=2\frac{\lambda}{\pi}V_s
\ee 
with LL accuracy. We obtain 
(with this precision $V_s=E_s$)
\be
E_s=V_s=-\frac{\lambda^{1+2\frac{\lambda}{\pi}}}{r}
\,.
\label{VsRGLL}
\ee
This result is correct with $\lambda^{n+1}\ln^n \lambda $ precision and is, 
obviously, gauge independent.  

We could consider the naive extrapolation to strong coupling of the above result. 
It does not agree with the AdS/CFT conjecture. 
We obtain $\lambda^{2\frac{\lambda}{\pi}}$ versus the expected 
$\sqrt{\lambda}$. This is not surprising as we have no reasons to believe 
our result should hold at strong coupling, the opposite limit with respect to which 
our result has been computed. 

We could consider playing a little bit and see how higher orders in 
the multipole expansion affect the result obtained in Eq. (\ref{VsRGLL}). Therefore, 
we consider the following RG equation\footnote{
Note that this result should only be considered for illustrative purposes, 
since contributions potentially of the same order (or even more 
important) have not been included in the anomalous dimensions.}
\be
\nu \frac{d}{d\nu}V_s=2\frac{\lambda}{\pi}V_s+\frac{2}{3}\frac{\lambda}{\pi}r^2V_s^3
\,.
\ee 
We obtain
\be
V_s
=
-\frac{\lambda^{1+2\frac{\lambda}{\pi}}}{r}\frac{1}{\sqrt{1+\frac{\lambda^2}{3}
(1-\lambda^{4\frac{\lambda}{\pi}})}}
\,.
\ee
For this result is not possible to take the strong coupling limit, since it becomes imaginary.

We would like to finish this section with some remarks. 
The order at which infrared logarithms appear in the static potential is 
easily visualized in the effective theory. In the case of QCD infrared logarithms first appear at 
${\cal O}(\als^4/r)$. This follows from the multipole expansion suppression of the interaction 
of the octet and singlet field through the chromoelectric field. There is no such suppression in the 
supersymmetric case, as we can see from the third line of Eq. (\ref{pnrqcd0}), which 
explains the 
appearance of such effects already at ${\cal O}(\als^2/r)$. On the other hand these effects 
can not come from gluons alone, since it would correspond to the pure QCD case.  Therefore, the 
logarithms have to be associated with the scalars. This is seen quite clearly 
in the effective theory. 

\section{The static potential at one loop}

In this section we compute the singlet static potential and 
energy at one loop (as well as some other matching coefficients). 
This computation is necessary for an eventual complete NLL evaluation. 
Moreover, this example will allow us to visualize how the factorization 
between the soft and ultrasoft scale takes place in 
dimensional regularization including finite pieces. 

We first compute the soft piece, ie. the static potential.
Diagrams with self energy cancel with diagrams with internal vertices \cite{Erickson:2000af}. 
Therefore, only the ladder diagram (and the crossed one if we go beyond the large $N_c$) 
has to be considered. Note that this is only true in the Feynman gauge. 
In any case, the final result is gauge independent. Therefore, the expression of the bare 
static singlet potential in $D=4+2\epsilon$ dimensions in momentum space reads
\bea
{\tilde V}_{s,B}&=&-2C_Fg^2\frac{1}{{\bf k}^2}
\left\{1-C_Ag^2{\bf k}^{2\epsilon}
\frac{\Gamma[1-\epsilon]\Gamma^2[\epsilon]}
{2^{3+2\epsilon}\pi^{2+\epsilon}\Gamma[2\epsilon]}
\right.
\nn
\\
&&
\left.
+{\cal O}(g^4)\right\}
\,.
\eea
After subtraction of the divergent piece in the $\MS$ scheme 
one can take the $D \rightarrow 4$ limit in the potential. It reads 
\be
\label{tildeVMS}
{\tilde V}_{s,\MS}=-2C_Fg^2\frac{1}{{\bf k}^2}
\left\{1-2\frac{\lambda}{\pi}\ln(k/\nu)+{\cal O}(\lambda^2)\right\}
\,.
\ee 
We see that there is no finite piece associated to the logarithm  
in the $\MS$ scheme. Eq. (\ref{tildeVMS}) is the initial condition 
of the singlet static potential for the eventual renormalization 
group equation at NLL. At NLL 
we would also need the initial conditions for $V_{\Phi}$ and 
$V_A$. Using arguments analogous to those in Ref. \cite{Brambilla:2006wp} we can 
conclude that there are no one loop corrections to those matching coefficients 
and 
\be
V_{\Phi}=1+{\cal O}(\lambda^2)
\,,
\ee
\be
V_{\Phi_O}=1+{\cal O}(\lambda^2)
\,,
\ee
\be
V_{A}=1+{\cal O}(\lambda^2)
\,.
\ee
The bare potential in position space reads
\bea
&&V_{s,B}=-\frac{2C_Fg^2r^{-2\epsilon}}{4\pi}\frac{1}{r}\frac{\Gamma[1/2+\epsilon]}{\pi^{1/2+\epsilon}}
\bigg\{1
\\
&&
\qquad
\left.
-C_Ag^2r^{-2\epsilon}
\frac{\Gamma[1/2+2\epsilon]\Gamma^2[\epsilon]}
{2^{3}\pi^{2+\epsilon}\Gamma[2\epsilon]\Gamma[1/2+\epsilon]}
+{\cal O}(g^4)\right\}
\,.
\nn
\eea
In the $\MS$ it reads
\bea
&&
V_{s,\MS}=-\frac{2C_Fg^2r^{-2\epsilon}}{4\pi}\frac{1}{r}\frac{\Gamma[1/2+\epsilon]}
{\pi^{1/2+\epsilon}}
\bigg\{
1
\\
&&
\nn
\qquad
-C_Ag^2r^{-2\epsilon}
\frac{\Gamma[1/2+2\epsilon]\Gamma^2[\epsilon]}
{2^{3}\pi^{2+\epsilon}\Gamma[2\epsilon]\Gamma[1/2+\epsilon]}
\\
&&
\qquad
\left.
+\frac{C_Ag^2{\nu}^{2\epsilon}}{4\pi^2}
\left(\frac{1}{\epsilon}+\gamma_E-\ln(4\pi)\right)
+{\cal O}(g^4)\right\}
\,,
\nn
\eea
which in four dimensions reduces to
\be
\label{VMS}
V_{s,\MS}=-\frac{2C_F\als}{r}\left\{1+2\frac{\lambda}{\pi}
\left[\ln(\nu r)+\gamma_E\right]+{\cal O}(\lambda^2)\right\}
\,.
\ee
Eq. (\ref{VMS}) is the relevant object to be introduced in a Schroedinger 
equation. In order to obtain $E_s(r)$ we also need the ultrasoft contribution, 
ie. we need to compute the diagram in Fig. \ref{figus} (with the $V_{\Phi}$ vertex) including finite pieces. 
From this computation we obtain the following correction to the singlet static energy
\bea
\delta V^{US}_{s,B}
&=&
2C_Fg^2(V_o-V_s)^{1+2\epsilon}
\frac{\Gamma[3+2\epsilon]\Gamma[-2-2\epsilon]}
{2^{2+2\epsilon}\pi^{3/2+\epsilon}\Gamma[3/2+\epsilon]}
\nn
\\
&&
+{\cal O}(g^4)
\,.
\eea
Finally, the energy of two static sources in the fundamental and anti-fundamental 
representation in ${\cal N}=4$ gluodynamics reads
\be
E_s(r)=
V_{s,B}+\delta V^{US}_{s,B}
=
V_{s,\MS}+\delta V^{US}_{s,\MS}
\,,
\ee
which in four dimensions reduces to the following expression
\be
\label{EsNLO}
E_s(r)
=
-\frac{2C_F\als}{r}\left\{1+2\frac{\lambda}{\pi}
\left[\ln(2\lambda)+\gamma_E-1\right]+{\cal O}(\lambda^2)\right\}
\,.
\ee

\section{Ordinary Wilson loop and its associated static energy}

Besides the locally 1/2 BPS Wilson loop considered in the previous sections, one could 
also consider the ordinary Wilson loop, which we define by eliminating the 
interaction of the static sources with the scalars in Eq (2), i.e.
\be
W_C=\frac{1}{N_c}{\rm Tr} {\cal P} 
e^{-ig\int_C d\tau A_{\mu}{\dot x}^{\mu}}
\,.
\ee
One can then redo the analysis of the previous sections for this case. 
Here we would like to compute the leading logarithmic corrections to the 
static energy and potential. 
In order to do so we can use the effective Lagrangian in Eq. (\ref{pnrqcd0}) eliminating the 
terms proportional to the scalars fields\footnote{At present, we can not 
claim that those coefficients vanish at any order in $\als$ but, at most, they 
are ${\cal O}(\als^2)$. This implies that the contribution from those terms is suppressed 
compared with those due to the ${\rm Tr}\{{\rm O}{\bf E}{\rm S}\}$ term.}, rescaling by a 
factor 1/2 the static potentials in Eq. (\ref{Vstree}), and redefining $\lambda \equiv N_c\als/2$. 
This implies that, to the order of interest, 
$\gamma_1=0$ and $\gamma_3$ is 1/2 the value quoted in Eq. (\ref{g3}).  
From this exercise, we learn the 
important lesson that the leading $\ln \als$ correction to the ordinary Wilson 
loop static energy in ${\cal N}=4$ supersymmetric Yang-Mills theory 
is the same to the one obtained in QCD 
\cite{Brambilla:1999qa}. The difference with QCD is due to the fact that 
$\als$ itself does not run. This implies that there are no 
${\cal O}(\als^{n+3}\ln^n(\als))$ terms in the static singlet potential 
except for $n=1$ (unlike in QCD \cite{Pineda:2000gz}). These findings 
can be summarized in the following equation, which is correct with NNLL accuracy, 
\be
\label{Esordinary}
E_s(r)=-\frac{C_F\als}{r}
\left(
1+a_1\als+a_2\als^2+\frac{C_A^3}{12}\frac{\als^3}{\pi}\ln(C_A\als)
\right)
\,.
\ee
The coefficient $a_1=N_c/\pi$ has been obtained in Ref. \cite{Alday:2007he}. 
We confirm this result. The coefficient $a_2$ is at present unknown.
 
For the static singlet potential one should replace $C_A\als$ by $\nu/r$ in the 
logarithmic term, where $\nu$ would correspond to the factorization scale. 

\section{Conclusions}

Eqs. (\ref{VsRGLL}), (\ref{EsNLO}) and (\ref{Esordinary}) 
are the main results of our paper. 

We have obtained the singlet 
static energy (and potential) with LL and NLO accuracy for the 
1/2 BPS static Wilson loop. We have provided with 
expressions at arbitrary dimensions, as well as with the formalism (based on effective 
field theories), that could be relevant for future computations.  
We have illustrated how the merge of the soft and ultrasoft contribution takes 
place in the case of the singlet static energy including finite pieces. This may be 
of help in order to visualize how things will work in the case of QCD, where this mixing 
takes place at three loops. 

A naive interpolation to strong coupling of the renormalization group improved LL result 
does not agree with the supergravity conjecture.

For the ordinary Wilson we have computed the 
singlet static energy and potential with NLO accuracy and we have also performed 
the resummation of logarithms with NNLL accuracy up to the initial matching condition  
$a_2$.

\medskip

I acknowledge discussions with B. Fiol, J. Gomis, A. Vairo and J.H. Kuhn,  
who somewhat triggered this investigation by asking me what the effect of a massless Higgs would 
be in the static potential. This work is partially supported by the 
network Flavianet MRTN-CT-2006-035482, by the spanish 
grant FPA2007-60275, by the catalan grant SGR2005-00916 and by a
{\it Distinci\' o} from the {\it Generalitat de Catalunya}.


\begin{references}

\bibitem{AdS}
  J.~M.~Maldacena,
  Adv.\ Theor.\ Math.\ Phys.\  {\bf 2}, 231 (1998)
  [Int.\ J.\ Theor.\ Phys.\  {\bf 38}, 1113 (1999)].

\bibitem{Rey:1998ik}
  S.~J.~Rey and J.~T.~Yee,
  Eur.\ Phys.\ J.\  C {\bf 22}, 379 (2001)

\bibitem{Vsstrong}
  J.~M.~Maldacena,
  Phys.\ Rev.\ Lett.\  {\bf 80}, 4859 (1998).

\bibitem{Erickson:1999qv}
  J.~K.~Erickson, G.~W.~Semenoff, R.~J.~Szabo and K.~Zarembo,
  Phys.\ Rev.\  D {\bf 61}, 105006 (2000).

\bibitem{Erickson:2000af}
  J.~K.~Erickson, G.~W.~Semenoff and K.~Zarembo,
  Nucl.\ Phys.\  B {\bf 582}, 155 (2000).

\bibitem{Appelquist:1977es}
  T.~Appelquist, M.~Dine and I.~J.~Muzinich,
  Phys.\ Rev.\  D {\bf 17}, 2074 (1978).
  
\bibitem{Brambilla:1999qa}
  N.~Brambilla, A.~Pineda, J.~Soto and A.~Vairo,
  Phys.\ Rev.\  D {\bf 60}, 091502 (1999).

\bibitem{Drukker:1999zq}
  N.~Drukker, D.~J.~Gross and H.~Ooguri,
  Phys.\ Rev.\  D {\bf 60}, 125006 (1999).
  
\bibitem{Gomis:2006sb}
  J.~Gomis and F.~Passerini,
  JHEP {\bf 0608}, 074 (2006).
  
\bibitem{Kniehl:1999ud}
  B.~A.~Kniehl and A.~A.~Penin,
  Nucl.\ Phys.\  B {\bf 563}, 200 (1999).

\bibitem{Pineda:1997bj}
  A.~Pineda and J.~Soto,
  Nucl.\ Phys.\ Proc.\ Suppl.\  {\bf 64}, 428 (1998).
  
\bibitem{Brambilla:2004jw}
  N.~Brambilla, A.~Pineda, J.~Soto and A.~Vairo,
  Rev.\ Mod.\ Phys.\  {\bf 77}, 1423 (2005).

\bibitem{Brambilla:2003nt}
  N.~Brambilla, D.~Gromes and A.~Vairo,
  Phys.\ Lett.\  B {\bf 576}, 314 (2003).

\bibitem{Pineda:2000gz}
  A.~Pineda and J.~Soto,
  Phys.\ Lett.\  B {\bf 495}, 323 (2000).
  
\bibitem{Brambilla:2006wp}
  N.~Brambilla, X.~Garcia i Tormo, J.~Soto and A.~Vairo,
  Phys.\ Lett.\  B {\bf 647}, 185 (2007).
  
\bibitem{Alday:2007he}
  L.~F.~Alday and J.~Maldacena,
  arXiv:0710.1060 [hep-th].





\end{references}
\end{document}